\documentclass[conference]{IEEEtran}

\usepackage{xspace}
\usepackage{etoolbox}
\makeatletter

\patchcmd{\@makecaption}
  {\scshape}
  {}
  {}
  {}
\makeatletter
\patchcmd{\@makecaption}
  {\\}
  {.\ }
  {}
  {}
\makeatother
\def\tablename{Table}

\usepackage{amsmath,amsfonts}
\usepackage{algorithmic}
\usepackage{graphicx}

\usepackage{textcomp}
\usepackage{todonotes}
\usepackage{booktabs}
\usepackage{tabularx}
\usepackage{multirow}
\usepackage{arydshln}

\usepackage{xurl}
\usepackage{balance}
\usepackage{float}
\usepackage{stfloats}
\usepackage{tikz}
\usepackage{xcolor}
\usepackage{bbding} 
\usepackage{threeparttable} 
\usepackage{makecell}

\usepackage{array}
\newcolumntype{L}[1]{>{\raggedright\let\newline\\\arraybackslash\hspace{0pt}}m{#1}}
\newcolumntype{C}[1]{>{\centering\let\newline\\\arraybackslash\hspace{0pt}}m{#1}}
\newcolumntype{R}[1]{>{\raggedleft\let\newline\\\arraybackslash\hspace{0pt}}m{#1}}

\DeclareRobustCommand*{\circled}[1]{%
  \tikz[baseline=(char.base)]{%
    \node[shape=circle,draw,fill=white,inner sep=0.7pt,
          minimum size=1.15em,text=black,font=\scriptsize\bfseries] (char) {#1};}}
\newcommand{\flowstep}[1]{\unskip~(\circled{#1})}

\def\BibTeX{{\rm B\kern-.05em{\sc i\kern-.025em b}\kern-.08em
    T\kern-.1667em\lower.7ex\hbox{E}\kern-.125emX}}

\definecolor{brickred}{rgb}{0.8, 0.25, 0.33}

\newif\ifshowreviewnotes
\showreviewnotesfalse
\ifshowreviewnotes
  \providecommand\wenhao[1]{\textcolor{blue}{\{\textbf{wenhao:} {\em#1}\}}}
  \providecommand\luowu[1]{\textcolor{blue}{\{\textbf{luowu:} {\em#1}\}}}
  \providecommand\link[1]{\textcolor{orange}{\{\textbf{link:} {\em#1}\}}}
\else
  \providecommand\wenhao[1]{}
  \providecommand\luowu[1]{}
  \providecommand\link[1]{}
\fi

\newcommand\sysname{\textsc{NACRE}\xspace}

\newenvironment{packeditemize}{
\begin{list}{$\bullet$}{
\setlength{\labelwidth}{2pt} 
\setlength{\itemsep}{0pt}
\setlength{\leftmargin}{\labelwidth}
\addtolength{\leftmargin}{\labelsep}
\setlength{\parindent}{0pt}
\setlength{\listparindent}{\parindent}
\setlength{\parsep}{1pt} 
\setlength{\topsep}{1pt}}}{\end{list}}

\DeclareRobustCommand*{\authorrefmark}[1]{\raisebox{0pt}[0pt][0pt]{\textsuperscript{\footnotesize\ensuremath{\ifcase#1\or *\or \dagger\or \ddagger\or%
    \mathsection\or \mathparagraph\or \|\or **\or \dagger\dagger%
    \or \ddagger\ddagger \else\textsuperscript{\expandafter\romannumeral#1}\fi}}}}
\usepackage{bbding} 

\definecolor{indianred}{rgb}{0.8, 0.36, 0.36}
\usepackage{hyperref}
\hypersetup{
    colorlinks = true,
    allcolors = magenta,
}

\usepackage{caption}
\usepackage{subcaption}

\newcommand{\para}[1]{\vspace{3pt}\noindent\textit{\textbf{{#1. }}}}
\providecommand\ignore[1]{{}}

\providecommand\revision[1]{#1}

\usepackage{cite}
\usepackage{CJKutf8}
\begin{document}
\begin{CJK*}{UTF8}{gbsn}

\title{\LARGE \bf \sysname: Rethinking Confidential Containers through Native Architectural Support}

\author{
\IEEEauthorblockN{
Linke Song\IEEEauthorrefmark{1}\IEEEauthorrefmark{2},
Wenhao Wang\IEEEauthorrefmark{1}\IEEEauthorrefmark{2}\Envelope,
Weijie Liu\IEEEauthorrefmark{3},
Rui Hou\IEEEauthorrefmark{1}\IEEEauthorrefmark{2}}
\IEEEauthorblockA{\IEEEauthorrefmark{1}State Key Laboratory of Cyber Space Security, Defense Institute of Information Engineering, CAS, Beijing, China}
\IEEEauthorblockA{\IEEEauthorrefmark{2}School of Cybersecurity, University of Chinese Academy of Sciences, Beijing, China}
\IEEEauthorblockA{\IEEEauthorrefmark{3}Nankai University}
}

\maketitle

\begin{abstract}
Linux containers achieve high density and fast lifecycle operations by sharing the host
kernel, but this design also lets a compromised host inspect or modify container state.
Existing confidential-computing systems protect an enclave address space or an entire
guest operating system, while recent container-granularity systems still add a separate
protection context. These abstractions do not make a dynamic group of host-managed Linux
processes the architectural protection unit.

This paper presents \sysname, a RISC-V hardware-software co-design for native
confidential containers. Its key insight is to separate the host's authority to manage
resources from its authority to access or commit protected state. Hardware-recognized
container identities direct protected traps to an isolated S-mode agent, while an M-mode
monitor commits security-sensitive identity, mapping, and page transitions. The agent
delegates services to host Linux without changing \texttt{satp}; services that neither
access private bytes nor modify protected state also avoid M-mode. We prototype \sysname
by extending QEMU, OpenSBI, Linux, a trusted agent, and \texttt{runc}. The prototype
implements the single-container private-memory substrate and covered launch, fault,
fork/COW, user-access, and teardown paths. Across five lmbench syscall and pipe metrics,
the three-run means remain within 3.5\% of the \texttt{runc-origin} baseline. With the
eight nginx object-size means weighted equally, aggregate throughput is 1.9\% lower.
\end{abstract}

\section{Introduction}
\label{sec:introduction}

Linux containers have become a fundamental abstraction for deploying modern workloads. By combining namespaces, control groups, and filesystem isolation while sharing the host kernel, containers avoid virtual hardware and a separate guest operating system, thereby providing low startup latency, a small memory footprint, and efficient resource sharing~\cite{merkel2014docker,oci2025runtime}. These properties have made containers widely used in cloud-native services, edge platforms, and high-performance computing environments~\cite{cncf2021edge,cncf2026survey,kurtzer2017singularity}. More recently, AI-agent systems have begun launching containers on demand as sandboxes for tool invocation and generated-code execution. For example, LangGraph deploys self-hosted agent servers through container images and orchestration platforms~\cite{langgraph2026standalone}, while OpenHands isolates agent-generated commands in Docker containers~\cite{openhands2026docker}. These emerging workloads further increase the demand for execution environments that combine strong isolation with the efficiency and flexibility of native containers.

The efficiency of containers, however, relies on a strong security assumption: the host operating system is trusted. Namespaces and control groups restrict the visibility and resource usage of container processes, but their enforcement ultimately depends on the host kernel. Once the host kernel or a privileged administrator is compromised, the attacker can inspect container memory, modify page mappings and execution state, forge kernel-maintained container metadata, or access sensitive data such as credentials and user inputs. Consequently, conventional containers cannot protect workloads deployed on infrastructure whose privileged software is not fully trusted.

In this work, we define a \emph{native confidential container} as a hardware-protected, dynamically managed group of Linux processes that continues to use host-native scheduling and resource management without relying on the host OS for the confidentiality or integrity of its execution state. The host may decide when processes run and which physical resources are allocated, but it must not be able to access protected container state, forge container membership, or convert resource-management authority into permission to read or modify protected resources. 
Realizing this abstraction requires three properties. First, the container must itself be the protection and lifecycle unit. Processes should be able to join and leave a protected container through operations such as \texttt{fork}, \texttt{clone}, \texttt{exec}, and \texttt{exit}, without reconstructing the protection boundary. Second, applications must retain native Linux process semantics, including copy-on-write memory, shared memory, signals, and synchronization among mutually trusting processes, without requiring source-code modification or a replacement programming model. Third, the host must continue to perform ordinary scheduling and resource-management functions, while access control and security-sensitive state transitions are enforced independently of the untrusted host. In particular, frequent container operations should not require switching into a VM, changing to a separate protected address space, or repeatedly invoking a high-privilege security monitor.

Existing confidential-computing abstractions do not simultaneously provide these properties. Enclave-based systems protect an isolated address space or a designated protected memory region~\cite{arnautov2016scone,gramine203255,shen2020occlum,lee2020keystone,feng2021penglai}. Enclave runtimes can improve application compatibility, but the protection boundary remains different from a dynamically changing container process group. Supporting multi-process applications therefore requires reconstructing Linux process abstractions inside a library OS or runtime, or coordinating process and resource sharing across protected execution contexts. This complicates dynamic process management and weakens compatibility with native Linux semantics.

Confidential virtual machines take the opposite approach by protecting an entire guest OS using hardware mechanisms such as AMD SEV, Intel TDX, and Arm CCA~\cite{kaplan2016amd,sev2020strengthening,tdx2020,cca2023}. VM-backed container systems, including Kata Containers and Confidential Containers, preserve Linux compatibility by placing containers together with a guest kernel and supporting services inside a protected VM~\cite{randazzo2019kata,confidentialcontainers,coco2026design}. However, the VM rather than the container becomes the hardware protection and resource-management unit. Container creation, memory allocation, process management, and I/O are consequently distributed across the guest kernel, virtual machine monitor, and host. The additional guest software stack also reduces the density and responsiveness that motivate native containers.

Recent container-granularity approaches attempt to bridge the gap between enclaves and CVMs by moving protection closer to container processes while retaining selected host services~\cite{hua2021tzcontainer,van2022blackbox,xu2024conmonitor,zhou2025rcontainer,song2026fasco}.
However, these systems still introduce additional protection contexts or virtualization layers to mediate host interactions.
As a result, they either sacrifice native Linux semantics or incur non-negligible overhead in frequent container operations.
The container itself is not a hardware-recognized execution identity that can directly govern process membership, resource ownership, and security-event delivery while coexisting with the host's native resource-management path.

The fundamental difficulty is that conventional hardware conflates two roles of the host OS: managing resources and serving as the security authority for those resources. Native containers require the host to retain the former role for efficiency, but an untrusted host cannot safely retain the latter. This tension is particularly visible on interrupts and exceptions. Delivering a confidential container's events directly to the host exposes its architectural state to untrusted privileged software. Delivering every event to a high-privilege security monitor avoids this exposure, but places frequent system interactions on a costly path involving additional privilege transitions and protection-context changes. Native confidential containers therefore require both a hardware-recognized container identity and a protected mediation path that does not place a high-privilege monitor on every common operation.

Based on this observation, we present \sysname, a RISC-V hardware-software co-design architecture for native confidential containers.
\sysname introduces \emph{container identity} as architectural security state. Hardware associates execution contexts and protected resources with a container identity and performs access checks and event routing using this identity rather than host-maintained process metadata. Trusted mechanisms control the creation, inheritance, and revocation of identities, preventing the host from forging container membership or obtaining access to protected resources through page remapping.

\sysname further separates frequent mediation from security-critical enforcement through two isolated components. A lightweight \emph{trusted container agent} executes at the same architectural privilege level as the host kernel but resides in a distinct hardware protection domain. Identity-aware event routing delivers interrupts and exceptions originating from confidential containers to this agent, preventing the untrusted host from becoming the first recipient of protected architectural state. The agent protects container context and mediates interactions with host-provided Linux services, while events from ordinary host processes continue to follow the conventional Linux path. A minimal \emph{security monitor} executes at a higher security level and handles only security-critical operations, such as container-identity management and protected-resource ownership transitions.

This separation allows frequent container-host interactions to remain in the native address-space and privilege context. The delegated trap path does not require entering a VM, switching page tables, or flushing the TLB; when a service neither accesses private bytes nor changes protected state, it also avoids the high-privilege monitor. The host can therefore continue to provide scheduling, physical-memory allocation, and ordinary Linux services, while hardware-enforced container identity prevents it from directly accessing protected state or fabricating security-relevant ownership. Dynamic process operations and inter-process sharing are expressed as changes within the same container identity rather than as coordination across independent protection contexts.

We implement a functional prototype spanning QEMU, OpenSBI, Linux, a trusted
agent, and \texttt{runc}. It integrates registration and \texttt{exec}, demand
paging, protected PTE updates, eager \texttt{fork} into fresh child pages,
copy-on-write, typed kernel access to private user buffers, and teardown with
the ordinary Linux lifecycle. The evaluated scope is the single-container
private-memory substrate; cross-container ownership, protected shared objects,
and attestation remain design-level mechanisms in this version.

QEMU-based functional tests exercise the covered launch and memory paths, and a
dedicated parent/child workload validates the implemented fork transaction. For
five lmbench syscall and pipe metrics, the three-run means are within 3.5\% of
the same-daemon \texttt{runc-origin} arm. With the eight nginx object-size means
weighted equally, aggregate throughput is 1.9\% lower. These results establish
the feasibility of the native execution path and characterize its observed
end-to-end behavior in QEMU.

\para{Contributions}
This paper makes the following contributions:

\begin{packeditemize}
    \item We formulate the abstraction of native confidential containers and identify the fundamental conflict between untrusted-host isolation and host-native resource management. We derive the requirements for making a dynamic Linux process group the hardware protection and lifecycle unit while preserving native multi-process semantics.

    \item We design \sysname, a RISC-V hardware-software co-design architecture that introduces hardware-recognized container identities, identity-aware resource protection, and identity-aware event routing. \sysname separates a peer-privilege trusted container agent for frequent mediation from a high-privilege security monitor that commits security-sensitive state; delegated services that access neither private bytes nor protected state avoid heavyweight protection-context transitions.

    \item We implement a QEMU-based prototype across OpenSBI, Linux, a trusted agent, and \texttt{runc}. The prototype preserves the covered native launch, private-memory, \texttt{fork}/COW, and teardown paths, and our preliminary evaluation characterizes common service paths and container workloads.
\end{packeditemize}

\section{Background \& Motivations}
\label{sec:preliminaries}

\subsection{Native Container Execution Model}
\label{subsec:linux-containers}

Linux containers isolate workloads through namespaces, control groups, and filesystem views while sharing the host kernel~\cite{merkel2014docker,oci2025runtime}.
After the container runtime establishes its namespaces, resource limits, filesystem view, and initial process, its tasks execute as ordinary Linux processes scheduled and managed by the host kernel.
Accordingly, executing the same OCI image inside a protected VM or runtime does not by itself preserve the native container execution model.
In this work, \emph{native} means that host Linux remains responsible for ordinary process and resource management, while hardware and trusted software independently protect security-sensitive container state.
This model has the following three properties.

\para{Host-resident process lifecycle}
A container comprises a dynamically changing group of host-scheduled Linux processes rather than a single address space or a guest operating system.
Operations such as \texttt{fork}, \texttt{clone}, \texttt{exec}, and \texttt{exit} directly create, transform, and terminate processes within this group.
Preserving this model requires that container processes remain directly visible to the host scheduler and process management subsystem, rather than being recreated through an additional guest OS or isolated runtime layer.

\para{Host-resident resource management}
Container resources follow the same lifecycle as ordinary Linux processes.
For example, \texttt{brk} and \texttt{mmap} create and modify virtual memory regions, page faults trigger on-demand physical page allocation, and \texttt{fork} establishes copy-on-write relationships between parent and child processes.
Linux maintains page tables, reverse mappings, page cache states, memory accounting, and reclamation metadata throughout this lifecycle.
Thus, preserving native resource management means that resource allocation, mapping, and reclamation remain part of the host Linux execution path, while accesses to confidential container state are selectively protected from the untrusted host.

\para{Host-resident Linux services}
Containers rely on host Linux for system calls, page fault handling, signals, file I/O, and inter-process communication.
Processes in the same container use native mechanisms such as shared memory, futexes, pipes, and sockets rather than communicating through a guest kernel.
Therefore, ordinary services continue to be implemented by host Linux. A service that neither accesses private bytes nor changes protected state should not require entering a guest OS, switching to a separate protected address space, or invoking a higher-privilege security monitor.

\subsection{Mismatch Between Existing Works and Native Containers}
\label{subsec:tee-container-mismatch}

\begin{table*}[t]
\centering
\small
\setlength{\tabcolsep}{3pt}
\renewcommand{\arraystretch}{1.2}
\caption{Comparison of protection units and Linux service paths in confidential-container approaches.}
\label{tab:tee-container-mismatch}
\begin{tabular}{@{}C{0.15\textwidth}C{0.17\textwidth}C{0.20\textwidth}C{0.20\textwidth}C{0.22\textwidth}@{}}
\toprule
\textbf{Approach} &
\textbf{Protection Unit} &
\textbf{Process and Resource Lifecycle} &
\textbf{Linux Service Provider} &
\textbf{Common Service Path} 
\\
\midrule
Enclave-based
\cite{arnautov2016scone,gramine203255,shen2020occlum,lee2020keystone,feng2021penglai}
&
Enclave / address space
&
Enclave runtime and host OS
&
Library OS, enclave runtime, and/or host OS
&
In-enclave handling or enclave--host transition
\\
CVM-based
\cite{sev2020strengthening,tdx2020,cca2023,randazzo2019kata,confidentialcontainers,coco2026design}
&
Confidential VM
&
Guest OS and VMM/host
&
Guest kernel
&
Guest-kernel path; VMM/host involvement for virtualized resources and I/O
\\
TEE-in-Container
\cite{lu2026arca}
&
Backend-specific enclave or CVM
&
OCI runtime outside the TEE; workload runtime inside
&
Backend dependent
&
Backend-specific enclave or VM path
\\
BlackBox
\cite{van2022blackbox}
&
Protected physical address space
&
Host Linux under a confidential security monitor
&
Host Linux through that monitor
&
EL2 interposition and protected-address-space switching
\\
ConMonitor
\cite{xu2024conmonitor}
&
Per-container nested-page-table context
&
Host Linux under ConVisor control
&
Host Linux and Container Guardian
&
VMFUNC-assisted nested-page-table switching
\\
RContainer
\cite{zhou2025rcontainer}
&
Container and con-shim protected by guarded page tables
&
Deprivileged host Linux and trusted mini-OS
&
Host Linux through mini-OS
&
Mini-OS mediation with EL3-assisted page-table switching
\\
Fasco
\cite{song2026fasco}
&
Container Realm
&
System Realm and Realm Management Monitor
&
System Realm and Trim OS
&
Monitor-mediated Realm transition and shared buffers
\\
\sysname
&
Container identity
&
Host Linux under identity-based protection
&
Host Linux through trusted agent
&
CID-directed S-mode handoff; M-mode only for private data or protected commits
\\
\bottomrule
\end{tabular}
\end{table*}

Existing confidential-computing systems map container workloads onto different hardware protection units.
The resulting systems differ not only in isolation granularity, but also in which software manages process and resource lifecycles and which execution path is taken when a protected workload requests Linux services.
Table~\ref{tab:tee-container-mismatch} summarizes these differences.

\para{Enclave-based approaches}
Enclave systems protect an application address space or a designated memory region.
Systems such as SCONE, Gramine, Occlum, Keystone, and Penglai provide Linux functionality through enclave runtimes, library OSes, or protected host interfaces~\cite{arnautov2016scone,gramine203255,shen2020occlum,lee2020keystone,feng2021penglai}.
Their hardware protection unit, however, remains an enclave rather than a dynamically changing group of host-managed Linux processes.
Multi-process abstractions and resource-sharing semantics must consequently be implemented inside the enclave runtime or coordinated across the enclave boundary.
Although these systems can execute containerized applications, their process and resource lifecycles do not directly coincide with those of native host containers.

\para{CVM-based approaches}
Confidential VM (CVM) technologies such as AMD SEV, Intel TDX, and Arm CCA protect a guest execution environment from an untrusted host~\cite{sev2020strengthening,tdx2020,cca2023}.
Kata Containers and Confidential Containers use this protection by placing a container or pod, together with a guest kernel and supporting services, inside a CVM~\cite{randazzo2019kata,confidentialcontainers,coco2026design}.
Ordinary application system calls may be handled entirely by the guest kernel and therefore do not necessarily cause VM exits.
Nevertheless, process creation, virtual-memory management, and Linux resource accounting occur inside the guest, while physical-resource allocation and virtualized I/O remain coordinated with the VMM and host.
The CVM, rather than the host-visible container process group, is therefore the hardware protection and resource-management boundary.

Arca explores a different TEE-in-Container organization that keeps orchestration logic outside the trusted execution environment and instantiates an independent TEE for each workload~\cite{lu2026arca}.
Since Arca supports SGX, TDX, and SEV backends, its concrete protection unit and service path remain determined by the selected enclave or CVM mechanism.

\para{Container-granularity approaches}
Recent approaches move the protection boundary closer to containers by protecting container processes or runtimes while retaining selected host services.
TZ-Container, BlackBox, ConMonitor, RContainer, and Fasco explore different combinations of trusted execution environments, virtualization extensions, and container-level protection mechanisms~\cite{hua2021tzcontainer,van2022blackbox,xu2024conmonitor,zhou2025rcontainer,song2026fasco}.
Despite their different implementations, these approaches introduce a separate protection context or mediation layer between containers and the host OS.
As a result, frequent container operations may require transitions across nested page tables, protected memory contexts, or trusted execution layers.
These additional transitions break the direct execution path between container processes and host Linux, complicating native resource management and increasing the overhead of frequent container operations.

\subsection{RISC-V Support and Missing Primitives}
\label{subsec:riscv-preliminaries}

On a conventional non-virtualized RISC-V system, applications execute in U-mode, Linux executes in S-mode, and platform firmware or a security monitor executes in M-mode.
M-mode may delegate selected exceptions and interrupts to S-mode through the \texttt{medeleg} and \texttt{mideleg} registers.
Once a trap is delivered to S-mode, \texttt{stvec} determines the S-mode trap-vector base address.
The \texttt{satp} register selects the root page table, address-space identifier, and virtual-memory translation mode.
S-mode software may request services from an M-mode execution environment through the Supervisor Binary Interface (SBI)~\cite{riscvprivileged,riscvsbi}.

These mechanisms separate software by privilege level or virtual address space, but they do not directly represent a protected container identity.
Placing all mediation logic in M-mode would cause frequent container events to enter the highest privilege level and would enlarge the highly privileged trusted computing base.
Placing the mediator in U-mode would leave its execution and memory mappings under the control of an untrusted S-mode kernel.
A native confidential container instead requires trusted mediation at S-mode efficiency while preventing host Linux, which also executes in S-mode, from accessing or modifying the mediator.

\para{Same-level protection-domain isolation}
Physical Memory Protection (PMP) allows M-mode to define physical-address permissions for less-privileged software~\cite{riscvprivileged}.
However, standard PMP checks depend on privilege level and physical address.
They do not distinguish two software components that both execute in S-mode.
PMP can therefore isolate M-mode from S-mode, but it cannot continuously distinguish trusted S-mode agent accesses from untrusted S-mode Linux accesses without reconfiguration by higher-privilege software.
Native confidential containers require an additional hardware-recognized execution identity that applies different access permissions to same-privilege components.

\para{Container-identity-aware trap routing}
Standard RISC-V trap delegation selects a destination privilege level according to the event type, and \texttt{stvec} provides one active S-mode trap-vector configuration for a hart.
Neither mechanism selects a handler according to the identity of the U-mode process that caused the event.
Consequently, traps from an ordinary host process and a confidential-container process follow the same S-mode entry path.
Routing all such traps first through M-mode would provide trusted dispatch but would place M-mode on every common event path.
Native confidential containers instead require hardware to route a trap according to both its cause and the container identity of its source.

\para{Lightweight same-level handoff}
Separating the trusted agent from host Linux with different address spaces would ordinarily require changing \texttt{satp}.
Although writing \texttt{satp} does not itself mandate a complete TLB flush, changing address spaces and reusing translation state may require \texttt{SFENCE.VMA} operations and associated TLB synchronization.
More importantly, address-space switching would place a new translation context on every mediated service request.
Entering M-mode to change protection state would introduce an additional privilege transition.
The required mechanism should instead transfer control between the trusted agent and host Linux while retaining the current address-translation context, and avoid M-mode when the delegated service accesses neither private bytes nor protected state.

\section{Threat Model}
\label{sec:threatmodel}

We consider each confidential container an independent protection domain
identified by a \texttt{cid}; processes and threads within the same
\texttt{cid} share a trust boundary. Protected state includes private memory,
register contexts, protected page tables, task/address-space bindings, and the
security metadata maintained by the trusted agent and security monitor. Data
explicitly released to an untrusted service or device leaves this boundary
unless an end-to-end protocol protects it.

The adversary controls host Linux, container runtimes, management software, and
drivers, and may execute arbitrary S-mode code. It can manipulate scheduling,
memory allocation, candidate page tables, traps, identifiers, and host-service
results, and it may run malicious containers. Host-provided metadata, mappings,
pointers, and resource states are therefore untrusted proposals: they cannot by
themselves authorize a protected-state transition. The trusted agent mediates
container traps before delegating selected services to Linux, but this mediation
does not guarantee the correctness or availability of those services.

We trust the processor and \sysname architectural mechanisms, the OpenSBI
security monitor, the trusted agent, the platform boot chain, and the
cryptographic primitives used by deployment. Container owners are trusted to
provide valid images, keys, and policies, and code admitted to one \texttt{cid}
is mutually trusting. We assume authenticated boot establishes the monitor and
agent before Linux runs, and that an IOMMU or equivalent mechanism prevents
host-controlled DMA from reaching protected memory. A device-bound root key and
freshness mechanism are required for remote attestation. The QEMU prototype
assumes, rather than implements, these platform properties.

\sysname targets confidentiality and integrity of protected container state,
isolation among different \texttt{cid}s, and controlled mediation of
container--host interactions. It does not prevent denial of service, starvation,
or malicious termination, nor does it isolate mutually distrustful code placed
inside one \texttt{cid}. Physical attacks, hardware or cryptographic failures,
microarchitectural side channels, rollback-resistant persistent storage,
traffic analysis, and protection after plaintext is released for untrusted I/O
are outside scope. \S\ref{sec:security} separates this target contract from the
properties exercised by the current prototype.

\section{Design}
\label{sec:design}

\subsection{Design Overview}
\label{subsec:overview}

\para{Design scope}
This section presents the \sysname architecture and its security contract.
\S\ref{sec:impl} identifies the subset realized by the QEMU prototype.

\para{Design principles}
\sysname uses a confidential container, rather than an individual process, as
its protection domain. In the target architecture, a common \texttt{cid}
identifies the processes, threads, and authorized shared mappings within one
container, whereas a cross-container mapping requires explicit authorization.
The protection boundary therefore matches the Linux container abstraction
without turning every process into an independent confidential execution
environment.

\sysname preserves Linux's resource-management responsibilities without
granting Linux security authority. Host Linux continues to create and schedule
processes, allocate physical pages, resolve ordinary faults, and maintain VMAs,
reverse mappings, and resource accounting. Trusted components instead validate
and commit operations that affect confidential context, protected mappings,
page ownership, or sharing relationships. This division retains the native
container lifecycle while preventing Host Linux from obtaining confidential
state merely by modifying pagetables or kernel metadata.

Trusted mediation applies only where a security-sensitive transition requires
it. For frequent system calls and exceptions, a trusted agent first saves the
confidential context and exposes to Linux only the state needed to perform the
delegated host service. The M-mode monitor maintains the security state for
protected pagetables, execution bindings, and, in the target multi-container
design, page ownership and sharing. Ordinary processes and non-confidential
containers retain the direct Linux path. This separation avoids moving general
host services into M-mode and leaves the trap path of ordinary workloads
unchanged.

\begin{figure}[t]
    \centering
    \includegraphics[width=\columnwidth]{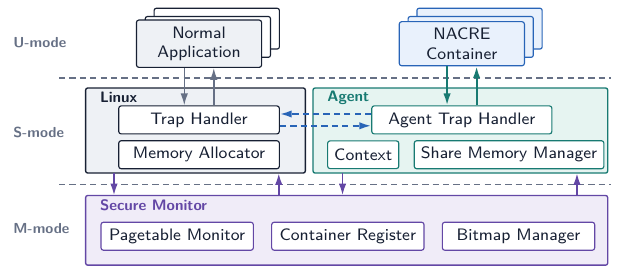}
    \caption{The \sysname architecture. A confidential container remains a
    U-mode Linux process, but its traps first enter an isolated S-mode trusted
    agent. The M-mode monitor maintains security state that neither S-mode
    component may modify directly.}
    \label{fig:nacre-overview}
\end{figure}

\para{Architecture}
\autoref{fig:nacre-overview} shows the three-layer \sysname architecture.
Ordinary applications and confidential containers both run as U-mode Linux
processes, but use different trap entries. Ordinary applications enter Linux
directly. A confidential-container trap first enters a trusted agent that, like
Host Linux, runs in S-mode. The agent saves registers and other confidential
context before Linux gains control, and determines which service state is
exposed to the host.

Although the trusted agent and Host Linux both execute in S-mode, they have
different authority. Linux schedules tasks, allocates memory, and provides
general file and network services. The agent handles confidential-container
traps, saves and restores confidential context, and prepares the state delegated
to Linux. In the target sharing design, it also maintains local policy state for
controlled shared objects. The agent cannot modify protected pagetables or
monitor metadata. It is part of the TCB, however, and the current access policy
permits agent-state execution to read or write private container data when
mediating a service. Linux cannot access the agent's code, data, or saved
context.

The M-mode monitor is the commit boundary for security state. It registers
container and task identities, maintains page roles, reference counts, and
protected pagetables, and validates operations that affect mappings or
execution identity. The multi-container design extends this metadata with page
ownership and sharing authorization. A delegated host service avoids M-mode
when it neither accesses private bytes nor changes protected state; typed
private-data operations and protected-state commits enter the monitor. The next
subsection describes the mechanisms that support this division of authority.

\subsection{Protection Substrate}
\label{subsec:memory}

The protection substrate combines execution identity with physical-page state.
Execution identity distinguishes an ordinary Linux task, the trusted-agent path
of a confidential container, and a delegated host-service path. Page state
determines whether Linux may access a physical page and which role that page
plays in a protected address space. Trusted hardware or the monitor maintains
both forms of state. Task flags, VMAs, and candidate PTEs supplied by Linux
describe requests but do not authorize them.

\begin{figure}[t]
    \centering
    \includegraphics[width=\columnwidth]{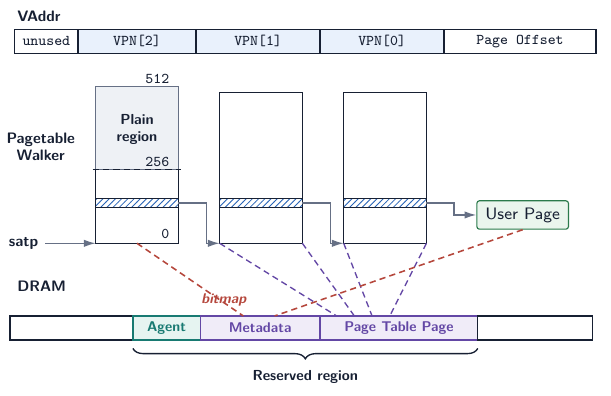}
    \caption{The \sysname memory-protection layout. Reserved memory contains
    the trusted agent, monitor metadata, and protected non-root pagetables.
    Address translation checks each mapping against the execution identity and
    physical-page role.}
    \label{fig:nacre-memory}
\end{figure}

\begin{table}[t]
\centering
\scriptsize
\setlength{\tabcolsep}{2pt}
\renewcommand{\arraystretch}{1.18}
\caption{Memory-protection metadata.}
\label{tab:memory-isolation-metadata}
\begin{subtable}[t]{0.45\columnwidth}
\centering
\caption{PMP permissions}
\label{tab:reserved-region}
\begin{tabularx}{\linewidth}{
>{\raggedright\arraybackslash}p{0.40\linewidth}
>{\raggedright\arraybackslash}X}
\toprule
\textbf{Object} & \textbf{Permission} \\
\midrule
Metadata & \textbf{M(RW)} \\
\mbox{Pagetable pages} & \textbf{M(RW)} $\parallel$ \textbf{SU(R)} \\
\bottomrule
\end{tabularx}
\end{subtable}\hfill
\begin{subtable}[t]{0.53\columnwidth}
\centering
\caption{Page roles}
\label{tab:bitmap-values}
\begin{tabularx}{\linewidth}{
>{\centering\arraybackslash}p{0.12\linewidth}
>{\raggedright\arraybackslash}p{0.17\linewidth}
>{\raggedright\arraybackslash}X}
\toprule
\textbf{Bits} & \textbf{Role} & \textbf{Meaning} \\
\midrule
00 & \texttt{NORM} & Ordinary Linux page \\
01 & \texttt{ROOT} & Root pagetable page named by \texttt{satp} \\
10 & \texttt{PRIV} & Confidential user-data page \\
11 & \texttt{PEND} & Confidential copy in a monitor transaction \\
\bottomrule
\end{tabularx}
\end{subtable}
\end{table}

\begin{figure}[b]
    \centering
    \includegraphics[width=0.82\columnwidth]{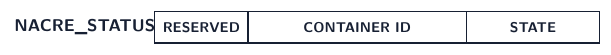}
    \caption{Logical execution state carried by \texttt{NACRE\_STATUS}.}
    \label{fig:nacre-status}
\end{figure}

\para{Trusted regions and execution identity}
As shown in \autoref{fig:nacre-memory}, reserved memory contains the trusted
agent, monitor metadata, and protected non-root pagetables, none of which Host
Linux may modify directly. Only M-mode may read and write monitor metadata, so
Linux cannot change page roles or private reference counts. S/U-mode
translation may read the protected non-root pagetable region, but only M-mode
may write it. PMP protects the agent region during boot. After container
initialization, the monitor-controlled \texttt{NACRE\_START\_AGENT} and
\texttt{NACRE\_END\_AGENT} registers delimit its physical range; hardware
combines these bounds with the current execution state to admit agent accesses
and reject ordinary S-mode accesses from Host Linux.

\texttt{NACRE\_STATUS} exposes the registered \texttt{cid} and current
control-flow state to the hardware, distinguishing trusted-agent execution from
delegated Linux execution. An unregistered task has a zero status and follows
the ordinary Linux access and trap rules. A registered task can change this
state only when the monitor restores its thread context or through the
controlled call and return instructions. In the target architecture, a
scheduler hint supplied by Linux therefore cannot open a confidential execution
window or grant Linux access to the agent or confidential pages. The prototype
boundary for this invariant is described in \S\ref{subsec:lifecycle}.

\para{Protected address spaces}
Linux continues to allocate the root pagetable page named by \texttt{satp} and
the user-data pages that back ELF segments and anonymous mappings. When these
pages enter the protected lifecycle, the monitor records them as
\texttt{ROOT} and \texttt{PRIV}, respectively. A \texttt{PRIV} page is
user-accessible only in a registered confidential-container execution state;
Host Linux cannot read or write it directly. Lower-level user pagetable pages
do not come from Linux. The monitor allocates them from the protected
pagetable region, whose PTEs hardware may read during translation but only the
monitor may update. \texttt{PEND} is the deny state for a confidential
destination while the monitor copies and installs it. Only M-mode can access
such a page. A successful copy transaction commits the destination PTE and
promotes the page to \texttt{PRIV}; an uncommitted destination must not reenter
the ordinary Linux lifecycle. The prototype's incomplete failure recovery is
discussed in \S\ref{sec:security}.

RISC-V Linux places a process's user address space in the lower half of the
canonical address space and shared kernel mappings in the upper
half~\cite{linuxriscvvm}. \sysname therefore protects only lower-half entries
of the root pagetable and leaves Linux in control of upper-half kernel
mappings. The monitor must commit all lower-half user PTE updates. Separating
\texttt{ROOT} from \texttt{PRIV} preserves Linux's control over kernel
mappings without allowing it to alter confidential user mappings. The current
prototype implements this rule for Sv39. The protection rule itself extends to
Sv48 and Sv57 by applying the same checks at their additional pagetable levels;
those modes are not implemented or evaluated in the current prototype.

\para{Translation and protected-state commit}
On a TLB miss, the current QEMU model selects the translation context from
\texttt{NACRE\_STATUS}. PMP and the agent-region bounds protect intermediate
pagetable-walk accesses, while the final physical address is additionally
checked against the page-role metadata. A translation enters the TLB only when
these checks permit the access. QEMU records a status-derived mode in the
translation context, so a later hit cannot be reused across an incompatible
mode. Controlled transitions between the trusted agent and Linux therefore do
not require a \texttt{satp} change or a wholesale TLB flush.

The monitor commits every change to protected page state or protected
pagetable permissions. Linux may perform the VMA lookup, classify the fault,
allocate a physical page, and prepare a candidate PTE. Before that proposal
becomes effective, the monitor checks it against the registered execution
context, root pagetable, destination slot, and applicable page-role policy. The
prototype mediates generic updates at this boundary; the typed fork-copy and
COW-replacement operations additionally combine copy and PTE publication into
one monitor transaction.

\subsection{Container Initialization and Runtime Mediation}
\label{subsec:lifecycle}

\begin{figure}[t]
    \centering
    \includegraphics[width=\columnwidth]{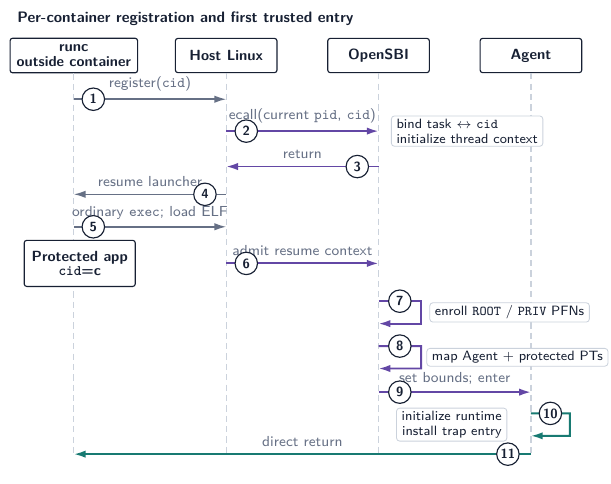}
    \caption{\sysname per-container runtime initialization and first agent entry
    after trusted setup. Linux follows the ordinary ELF loading path, while
    protected execution is admitted only after monitor validation and commit.}
    \label{fig:nacre-initial}
\end{figure}

\para{Container initialization}
\sysname separates trusted setup from per-container initialization. The target
deployment installs the monitor and agent into the reserved region in
\autoref{fig:nacre-memory} before Linux is treated as untrusted; the QEMU
prototype assumes this pretrusted initial state.
Initialization is the first container lifecycle transition: Linux-created
process state becomes
\sysname-protected state, but the process is still created by the ordinary
\texttt{runc} and Linux \texttt{exec} path.

\autoref{fig:nacre-initial} shows this runtime path. Before \texttt{exec},
\texttt{runc} issues a registration system call \flowstep{1}. Linux forwards
the request to the secure monitor. This registration is the initialization gate
for a Linux task to become a confidential-container task: the monitor records the
Linux task identifier (\texttt{pid}) and \sysname container identifier
(\texttt{cid}) in its registration state and creates the initial
per-thread \sysname context \flowstep{2}. The context is keyed by the registered
task identity and \texttt{cid}, and records the \texttt{NACRE\_STATUS} value
that the monitor will later install when the task is scheduled into the
confidential-container execution window. This value encodes the \texttt{cid}
and the control-flow location used by \sysname to distinguish the \texttt{LINUX}
state from the \texttt{AGENT} state. Tasks without a registered \sysname context
use the default zero value. This registration and thread-context state is
monitor-maintained state; it is separate from the per-page bitmap used to label
physical-page roles. The monitor then returns to Linux \flowstep{3}; Linux
resumes \texttt{runc} along the same path
\flowstep{4}. Registration does not map or execute the agent. This registration
call is the only change required to \texttt{runc} on the launch path; after it
returns, \texttt{runc} follows the ordinary OCI lifecycle and invokes the
standard Linux \texttt{exec} path. Later
\texttt{fork} and \texttt{clone} operations that create additional
confidential-container tasks must similarly create registered \sysname thread
contexts before those tasks can enter the confidential-container execution
window.
The QEMU prototype accepts the \texttt{pid} and \texttt{cid} supplied along
this cooperative launch path. It therefore validates lifecycle plumbing, not
resistance to forged or stale membership. A deployment must bind accepted
identities to monitor-controlled freshness and derive child membership from an
authorized parent transition; \S\ref{sec:security} treats this as an open link
in the current security argument.

\texttt{runc} later executes the container program and transfers
control to the post-\texttt{exec} application image. Linux handles the ordinary
\texttt{exec} path and builds the user address space from a prepared ELF image
\flowstep{5}. The current QEMU prototype treats this image as a pretrusted test
input. In the target deployment, an authenticated image-dispatch path must bind
the protected contents and public loading metadata to a manifest before the
monitor admits protected execution; that image-verification protocol is not
implemented in the prototype.

Control then enters the monitor with the application's user-mode resume context
\flowstep{6}. Before the agent is mapped, the monitor uses its page-monitor
logic to enroll the pages allocated during ELF loading into protected metadata:
\texttt{ROOT} for the root pagetable page and \texttt{PRIV} for
Linux-allocated user data pages \flowstep{7}. This enrollment also initializes
the private reference count for committed user-data mappings and rejects
physical frame numbers (PFNs)
that already carry an incompatible role. The security monitor then
updates the target process pagetable and maps the reserved agent
region into a user-space slot in the low-half part of the \sysname process
address space \flowstep{8}. This slot is chosen to avoid the ranges normally
occupied by the ELF image, heap, stack, and Linux-managed anonymous mappings, so
applications usually do not need binary or source changes. The agent mapping is
not part of the ELF image; it is installed only at this monitor-mediated mapping
step. Although the agent is mapped into a user-address-space range, the
pagetable pages carrying this mapping are not allocated or modified by Linux.
The monitor allocates them from the protected pagetable page area in
\autoref{fig:nacre-memory}, following the same monitor-only update rule used for
protected user mappings.
Before entering the mapped agent, the monitor switches protection from the
boot-time PMP guard to two monitor-controlled registers,
\texttt{NACRE\_START\_AGENT} and \texttt{NACRE\_END\_AGENT}, that Linux cannot
modify. These registers bound the agent physical region \flowstep{9}. Once this base-and-bound
guard~\cite{intel2026sdm} is enabled, M-mode and S-mode execution in the
\texttt{AGENT} state may access the region, whereas protected U-mode and Linux
S-mode accesses are rejected. The state transition, not S-mode privilege alone,
therefore distinguishes the agent from Linux.

The agent then initializes its own address-space memory and local runtime state
so it can later handle container requests \flowstep{10}. As part of this setup,
the agent records the virtual entries of the original Linux trap handlers used
for delegated services, and writes the virtual entry of its own trap handler
into \texttt{NACRE\_TWIN\_ENTRY}. The latter becomes the first-hop target for
later delegated traps from confidential-container execution. When this
completes, the agent jumps directly back to the user application using the
application resume context saved at \flowstep{6}, without re-entering the security
monitor \flowstep{11}.

\begin{figure}[t]
    \centering
    \includegraphics[width=\columnwidth]{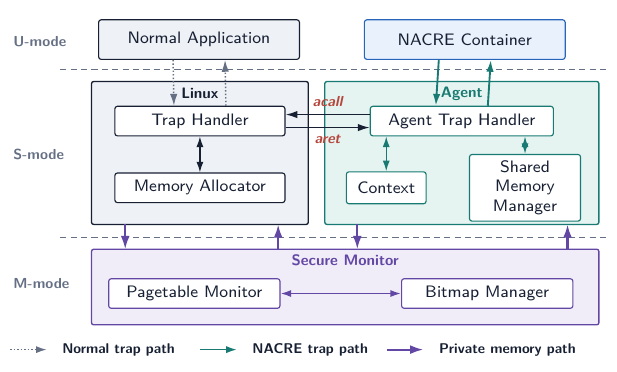}
    \caption{\sysname post-initialization runtime paths. Normal applications
    use the normal trap path, confidential-container traps use the \sysname trap
    path through the agent, and private-memory transitions enter the monitor
    when protected pagetable state, bitmap/refcount metadata, or confidential
    data-page copy/retirement must be committed.}
    \label{fig:nacre-procedure}
\end{figure}

\autoref{fig:nacre-procedure} summarizes the runtime control paths after
initialization. We first focus on the two trap paths: the normal trap path keeps
ordinary applications on the Linux trap handler, while the \sysname trap path
routes confidential-container traps through the agent before Linux executes the
delegated service.

\para{Scheduling window and thread binding}
The distinction between the first two paths in
\autoref{fig:nacre-procedure} is made by the execution window restored for the
scheduled task. After initialization, the monitor uses the registered \sysname
thread contexts to decide when a scheduled Linux task may enter the
confidential-container execution window. Linux still saves and restores ordinary
process context through its normal scheduler. The scheduling state specific to \sysname,
including the encoded \texttt{cid} and control-flow location in
\texttt{NACRE\_STATUS}, is recovered from the monitor-registered thread context
rather than trusted from Linux at context-switch time.

In the QEMU prototype, Linux marks candidate confidential-container tasks in
\texttt{task\_struct} and enters the monitor when a switch may involve one of
them. The monitor then saves the outgoing \sysname state or restores the
registered incoming state, including \texttt{NACRE\_STATUS}. This cooperative
path keeps ordinary context switches unchanged, but the Linux marker is not a
security authority: a malicious scheduler can omit the hook. The target
invariant is that a protected user return is accepted only for a fresh,
monitor-bound task, \texttt{cid}, and address-space context. The current
prototype does not yet provide the non-bypassable scheduling hook and freshness
mechanism needed to enforce that invariant against malicious Linux.

\para{Trap delegation and fast-path separation}
The normal and \sysname trap paths in \autoref{fig:nacre-procedure} differ only
when a registered confidential-container context is active. Normal tasks keep
the unmodified Linux path: hardware vectors their system calls and exceptions
through \texttt{stvec} to the Linux trap handler. A registered \sysname task is
distinguished by a nonzero \texttt{NACRE\_STATUS}. When such a task traps inside
the confidential-container execution window, the first hop goes to the agent
entry stored in \texttt{NACRE\_TWIN\_ENTRY} rather than directly to Linux.

The agent saves the protected register context and builds a delegated context in
Linux-readable memory. It then uses \texttt{scause} to select one of the Linux
trap-handler entries recorded during initialization and writes that handler
address into the
\texttt{NACRE\_TRAMPOLINE} register. Executing \texttt{acall} switches
\texttt{NACRE\_STATUS} to the \texttt{LINUX} state and transfers control through
that trampoline to the selected Linux handler. As part of this transfer, the
trampoline is changed to hold the agent return entry, so the later return has a
single controlled target. Linux then runs its ordinary handler over the delegated
context, without direct access to the protected container context.

When Linux finishes the service, its common return wrapper executes
\texttt{aret}. For a valid pending return in the \texttt{LINUX} state, the
instruction clears the pending state, transfers control through the trampoline,
and restores the \texttt{AGENT} state. The agent restores the Linux-produced
trap frame after clearing \texttt{SIE} and \texttt{SPP} and reseeding
\texttt{sscratch}; it does not validate the semantics of arbitrary host-service
results. Because the agent and Linux use the same process pagetable,
\texttt{acall} and \texttt{aret} do not change \texttt{satp}, switch
pagetables, or flush TLB entries. A delegated service also avoids M-mode when it
neither accesses private bytes nor changes protected mappings or metadata.
Private-buffer access instead uses the typed monitor operations described in
\S\ref{sec:impl}.

The result is a separation between the two trap paths: ordinary tasks trap
directly to Linux, while confidential-container traps return through the agent.
Monitor entries for private-byte access and protected-state changes are outside
the \texttt{acall}/\texttt{aret} fast path and use the private-memory path
described next.

\para{Core private-memory substrate}
The private-memory path in \autoref{fig:nacre-procedure} first establishes the
single-container substrate: monitor entries commit protected page-role or
mapping changes for one confidential container. The boundary is the protected
commit point. Linux proposes VM policy results through the ordinary kernel path:
it classifies faults, looks up VMAs, allocates backing pages, and maintains
\texttt{struct page}, reverse mapping, RSS, and LRU state. When the kernel path
reaches a result that would update a confidential mapping, copy confidential
contents, or retire protected state, control enters OpenSBI through an SBI ecall
carrying the operation type, target PTE location, and intended mapping. OpenSBI
uses the registered task and root context as validation inputs before it
performs the protected pagetable update or data-page operation. VMA flags and
Linux PTE contents route the request, but they are not security authority.

OpenSBI records per-page protected state in the page metadata shown in
\autoref{fig:nacre-memory}. The bitmap values in \autoref{tab:bitmap-values}
distinguish \texttt{NORM} pages, \texttt{PRIV} confidential data pages, and
\texttt{ROOT} root pagetable pages. Private reference counts are stored alongside
this per-page state and are writable only by OpenSBI. This counter is separate
from Linux's mapcount: it records the monitor-approved protected mappings that
currently hold each confidential PFN, and it is used to decide when a
\texttt{PRIV} PFN can be retired. The bitmap is therefore only a per-page role
encoding for the protected-memory substrate; it does not encode a confidential
container context, a thread context, or a complete ownership policy. Those
states are maintained separately by the monitor. More expressive shared-object
or multi-container policies can be layered on top of this substrate without
changing the basic rule that OpenSBI is the protected-state committer.

For a protected page fault, Linux performs VMA and fault resolution and obtains
the physical page together with its ordinary \texttt{struct page} and accounting
state. The PFN remains \texttt{NORM} at this point. Once the prepared leaf PTE
targets a protected user mapping, OpenSBI commits the transition. In the target
protocol, it verifies that the PTE slot belongs to the registered root pagetable
page or to a protected non-root pagetable page, checks that the target PFN can
enter the protected lifecycle for this registered address space, records the
PFN as \texttt{PRIV}, initializes or increments its private reference count,
and writes the PTE. The
root pagetable page and ordinary user data pages are still allocated by Linux;
OpenSBI protects them by assigning bitmap roles and mediating protected updates,
rather than by replacing Linux allocation. If the walk needs a new non-root
pagetable page, OpenSBI allocates it from the protected pagetable page area and
initializes the corresponding pagetable metadata before committing the leaf PTE.
The same path handles later demand faults after initialization and after
\texttt{exec} rebuilds an address space.

\para{Container lifecycle across processes}
\texttt{fork}, \texttt{exec}, COW, and exit extend the substrate from a single
protected address space to the normal Linux container lifecycle. Linux retains
its native \texttt{mm} duplication, pagetable traversal, VMA/rmap/RSS
accounting, and fault classification. The trusted agent brackets a distinct-mm
\texttt{fork} with a monitor-owned authorization window, and the monitor admits
at most one fresh child root for that operation. This relation prevents a
Linux-supplied root or process identifier from authorizing an unrelated
protected mapping.

Inherited private pages are copied eagerly rather than shared between the
parent and child. For each private parent leaf, Linux prepares a fresh child
destination and invokes \texttt{fork\_copy\_install}. The monitor validates the
parent root, selected child root, virtual address, source leaf, destination PFN,
and requested PTE shape. It changes the fresh destination from \texttt{NORM} to
\texttt{PEND} before copying the confidential bytes. Successful commit promotes
the destination to \texttt{PRIV} and publishes the child PTE atomically; no
child PTE is visible before that commit. The child therefore begins with private
copies while Linux still performs the surrounding native fork bookkeeping.

Linux also retains its ordinary writable-fault path for a protected leaf that
requires COW replacement. It allocates and accounts for the destination page,
then invokes the monitor's \texttt{cow\_replace} transaction. The monitor
validates the source and destination, changes the destination from
\texttt{NORM} to \texttt{PEND}, copies the confidential bytes, and atomically
replaces the protected PTE while promoting the destination to \texttt{PRIV} and
retiring the old protected reference. Both operations collapse what would
otherwise be a separable copy and PTE update into one monitor commit; they do
not rely on a Linux-controlled copy reservation.

The target recycle and retirement protocol closes the lifecycle. During \texttt{munmap}, COW
replacement, \texttt{exec} teardown, or process exit, Linux may still walk VMAs
and pagetables and reach the kernel sites that would normally clear PTEs.
OpenSBI validates each retired protected PTE, clears or replaces it, and
decrements the private reference count of the mapped PFN. When the count reaches
zero, OpenSBI first ensures that no protected PTE still holds the PFN, clears
the confidential contents, and only then returns the bitmap entry to
\texttt{NORM} so the PFN can reenter the ordinary Linux lifecycle.
During address-space destruction, batching is limited to protected non-root
pagetable pages whose protected leaf entries have already been retired. These
pages return to the protected pagetable page area rather than the Linux buddy
allocator. The root pagetable page is retired last by removing its \texttt{ROOT}
bitmap entry.

\para{Same-container controlled sharing}
Explicit shared-memory IPC is a separate layer above the private-memory and
multi-process lifecycle substrate. Eager private-page copying during
\texttt{fork} deliberately does not provide a shared-memory object model. In
the same-\texttt{cid} design, the agent acts as the policy and control plane for
approved shared objects. The agent may mediate object data, but it cannot commit
protected PTE or monitor-metadata changes. For each approved attachment,
OpenSBI records the object, \texttt{cid}, permissions, and live mapping before
installing the protected PTE. Detach revokes the mapping and stale translations
before the backing page can be reused. Linux therefore cannot turn a private PFN
into a shared object by writing PTEs alone.

\para{Cross-container ownership and isolation}
Cross-container isolation adds owner and generation metadata to protected
non-root pagetable pages and private PFNs. Every protected PTE update is bound to
the registered task, \texttt{cid}, root pagetable page, virtual address, target
PTE slot, and object grant, if any. A request to map, retire, or copy another
container's protected PFN is rejected unless a live monitor-visible grant names
the recipient and permissions. Revocation removes authorized mappings and stale
translations before the object generation can be reused.

\para{Membership transitions}
Container membership changes decide which tasks and address spaces may invoke
the private-memory transitions above. A \texttt{clone} that shares the address
space adds a task to the same \texttt{cid}, so the new thread remains inside the
existing protection domain only after the monitor records the new \sysname thread
context. A \texttt{fork} creates a new address space, so its root pagetable page and
inherited protected mappings must be derived through monitor-checked metadata
transitions. The child is not allowed to enter the confidential-container
execution window until its task identity, \texttt{cid}, and
\texttt{NACRE\_STATUS} value have also been registered as a \sysname thread
context. Futex synchronization then operates on already-approved shared pages
rather than defining a new protection boundary.

\subsection{Target Measurement and Attestation}
\label{subsec:attestation}

\sysname defines a measured container object over the monitor and agent
configuration, initial application image, accepted public mappings, and
container policy. Attestation binds this object to a fresh \texttt{cid} and
root-pagetable generation; later reports describe only state changed through
the protected transitions above. This lets a verifier reason about the Linux
container as one protection domain rather than attest a collection of unrelated
processes. Report generation and verifier integration are outside the current
prototype.

\section{Implementation}
\label{sec:impl}

The design section defines the protection rules and runtime paths. This section
describes where those rules are realized in the prototype. We focus on
implementation boundaries rather than restating the design: which components
were changed, which state is kept by Linux versus OpenSBI, and which Linux
events are converted into monitor-mediated commit points.

\subsection{Prototype Organization}

Our prototype combines modified QEMU 9.2.0, OpenSBI derived from v1.6, Linux,
a trusted agent, and a minimally changed \texttt{runc}. The RISC-V guest uses
an Ubuntu 22.04 root filesystem. QEMU serves as the architectural and functional
model; \S\ref{sec:evaluation} reports the corresponding experimental scope.

The prototype is organized to keep ordinary Linux container paths
intact. \texttt{runc} is modified only to issue one registration system call
before \texttt{exec}; the remaining OCI launch path is unchanged. Linux changes
are localized: ordinary \texttt{exec}, scheduling, VM policy, and kernel
metadata remain on their standard paths, while \sysname adds branch checks and
ecall exits at points that access private bytes or commit protected state. OpenSBI adds
SBI handlers that manage protected metadata, protected pagetable updates, and
registered context state. QEMU models the \sysname execution-state, twin-entry,
trampoline, and agent-bound registers; \texttt{acall}/\texttt{aret};
agent-region checks; and final-address page-role checks. The corresponding QEMU
fields are \texttt{nacc\_state},
\texttt{twin\_entry}, \texttt{trampoline}, \texttt{nacc\_sagent}, and
\texttt{nacc\_eagent}. The agent is a trusted runtime for trap delegation and
service preparation. Agent-state execution may access private container data,
but it cannot commit protected
pagetable or monitor-metadata changes; those operations remain within OpenSBI.

The rest of this section follows this component boundary. We first
describe boot-time agent installation and per-container registration, then
describe the monitor metadata, commit interfaces, private-memory transitions,
and modeled architectural support used by the prototype.

\subsection{Boot and Launch Integration}

\sysname builds a normal Linux boot artifact and appends two objects to it: the
agent payload and a descriptor page. The descriptor records the
location of the appended agent image and the reserved physical region where the
image must be installed. The QEMU prototype assumes that this artifact and the
initial container image are pretrusted inputs; it does not implement the target
authenticated-boot or image-manifest protocol.

OpenSBI performs the installation before transferring control to Linux. It
parses the descriptor, copies the agent payload into the reserved agent region
shown in \autoref{fig:nacre-memory}, clears both the appended agent copy and the
descriptor from the original boot artifact, and installs a temporary PMP rule
that makes the loaded agent region accessible only to M-mode. Linux can then
continue its ordinary boot path, but it cannot read or overwrite the installed
agent bytes.

Per-container launch therefore adds only one hook to the ordinary
container path. Before \texttt{exec}, \texttt{runc} issues the registration
system call described in \S\ref{subsec:lifecycle}, carrying the \sysname
container identifier (\texttt{cid}). Linux associates the request with the
current task and forwards the resulting \texttt{pid}/\texttt{cid} tuple to
OpenSBI. After the call returns, \texttt{runc} resumes
the ordinary OCI lifecycle, and Linux still performs the subsequent
\texttt{exec} and ELF loading work.

\subsection{NACRE SBI Commit Interfaces}

The implementation exposes the design's protected-state commit points through
NACRE-specific SBI ecalls. We use the paper-facing operation names in
\autoref{tab:nacre-sbi-interfaces}; the concrete Linux and OpenSBI trees group
these operations under their respective NaCC SBI extension identifiers. These
interfaces are not a general metadata
management API: they are the places where Linux, after following an ordinary
container or memory-management path, must enter OpenSBI before a protected
binding, pagetable update, or teardown action becomes effective.

\begin{table}[t]
\centering
\scriptsize
\setlength{\tabcolsep}{3pt}
\renewcommand{\arraystretch}{1.12}
\caption{Representative NACRE SBI operations (paper-facing names).}
\label{tab:nacre-sbi-interfaces}
\begin{tabularx}{\columnwidth}{@{}
>{\raggedright\arraybackslash}p{0.34\linewidth}
>{\raggedright\arraybackslash}X@{}}
\toprule
\textbf{Call} &
\textbf{Role in the protected lifecycle} \\
\midrule
\texttt{NACRE\_REGISTER} &
Records the initial \texttt{pid}/\texttt{cid} binding before
\texttt{exec}. \\
\texttt{NACRE\_INVOKE} &
Enters the agent path for the registered task and establishes the
runtime execution context. \\
\texttt{NACRE\_REEXEC} &
Reattaches after \texttt{exec}, when Linux has built a fresh
\texttt{mm} and root pagetable page. \\
\texttt{NACRE\_INVOKE\_CHILD} &
Invokes an already initialized child process through a shortened
invoke path. \\
\makecell[l]{\texttt{NACRE\_ATTACH}\\\texttt{\_FORKED\_CHILD}} &
Attaches a freshly forked child before its first protected
user-mode return, skipping the initial bootstrap. \\
\texttt{NACRE\_VERIFY} &
Saves or restores monitor-owned thread context at scheduler boundaries. \\
\texttt{NACRE\_TAG\_ROOT} &
Enrolls the current address-space root pagetable page for protected updates. \\
\texttt{NACRE\_REQ\_PTP} &
Allocates a protected non-root pagetable page from monitor-managed memory. \\
\texttt{NACRE\_SET\_PTES} &
Commits Linux-prepared PTE values after OpenSBI validates the target slots. \\
\makecell[l]{\texttt{FORK\_BEGIN}\texttt{/}\texttt{FORK\_END}} &
Opens and closes the monitor-owned authorization window around one
Agent-observed distinct-mm fork. \\
\makecell[l]{\texttt{FORK\_SELECT}\\\texttt{\_CHILD\_ROOT}} &
Selects at most one fresh child root for the active parent fork window. \\
\texttt{FORK\_COPY\_INSTALL} &
Copies one exact parent private leaf into a fresh child page and publishes the
child PTE as one monitor transaction. \\
\texttt{COW\_REPLACE} &
Copies and replaces an exact private leaf within one attached address space as
one monitor transaction. \\
\texttt{NACRE\_RETIRE\_ROOT} &
Retires state for the root pagetable page when the protected address space is replaced or
destroyed. \\
\texttt{NACRE\_RECLAIM\_PTP} &
Batch-reclaims protected non-root pagetable pages after leaf mappings are gone. \\
\texttt{NACRE\_UNREGISTER} &
Removes monitor-side task registration state during process exit. \\
\midrule
\texttt{NACRE\_AGENT\_MMAP} &
Maps the agent bootstrap region through OpenSBI during the first
agent entry. \\
\texttt{NACRE\_AGENT\_READY} &
Marks the first full agent initialization complete, enabling later
lightweight attaches. \\
\bottomrule
\end{tabularx}
\end{table}

\revision{\autoref{tab:nacre-sbi-interfaces} lists the representative
monitor entries used by the prototype. These calls appear at the points where
the native Linux lifecycle attaches a registered task, installs or updates a
protected PTE, or retires protected address-space state.}

\subsection{\texorpdfstring{\revision{Single-Container Lifecycle Integration}}{Single-Container Lifecycle Integration}}

\para{\revision{Registration and attachment}}
\revision{\sysname splits process identity registration from protected
address-space attachment. The registration call made by \texttt{runc} creates
or refreshes the monitor-side \texttt{pid}/\texttt{cid} record and per-task
execution context, but it does not yet attach a specific \texttt{mm}. After
\texttt{exec} succeeds, Linux has constructed the new address space and root
pagetable page, and the task enters OpenSBI through the invoke, reexec, or child
attach path. The first protected task performs the full agent bootstrap and
marks the agent runtime ready with the agent-side calls in
\autoref{tab:nacre-sbi-interfaces}. Later \texttt{exec} and child paths reuse
this installed runtime state: \texttt{NACRE\_REEXEC} refreshes the root and
trap-return context for the same registered task, while
\texttt{NACRE\_INVOKE\_CHILD} and
\texttt{NACRE\_ATTACH\_FORKED\_CHILD} are shortened variants for child
processes that avoid repeating the first-process bootstrap.}

\para{\revision{Protected PTE installation}}
\revision{Linux still owns the software memory-management state: VMAs,
\texttt{struct page}, RSS, reverse mappings, LRU state, and memcg accounting
remain on the native paths. The implementation changes the physical source and
write authority for protected pagetable pages. The root pagetable page is a
Linux-allocated page that OpenSBI tags as \texttt{ROOT}; non-root protected
pagetable pages are allocated from the monitor-managed PTP pool while Linux
keeps the ordinary page-table descriptors needed by its VM code. Instead of
adding separate security logic to every fault, fork, mprotect, and unmap branch,
\sysname redirects common RISC-V PTE primitives--install, write-protect,
exchange, and accessed-bit updates--when the target slot belongs to protected
pagetable state. Linux supplies the candidate PTE, target slot, virtual address,
and root context; OpenSBI validates the destination and PFN role, updates the
bitmap/reference-count metadata, and writes the protected PTE.}

\para{\revision{Fork and COW}}
\revision{\texttt{fork} continues to use Linux's native \texttt{mm} duplication,
page-table traversal, VMA/rmap/RSS accounting, and surrounding fault logic. At
the trusted trap entry, the agent recognizes a legacy \texttt{clone} with
\texttt{CLONE\_VM} clear and opens a monitor-owned fork window. Linux then
submits the root allocated by its normal \texttt{pgd\_alloc()} path. OpenSBI
admits at most one fresh child root for the window and records that exact
parent--child relation; Linux-supplied task or container identifiers remain
consistency inputs rather than authority. The child cannot enter its protected
execution window until its task context and selected root are attached.}

\revision{The implemented fork path does not publish the parent's private PFN into the child.
When native \texttt{copy\_page\_range()} reaches an inherited protected leaf,
Linux prepares a fresh destination page and invokes
\texttt{FORK\_COPY\_INSTALL}. Under the monitor lock, OpenSBI validates the
active fork window, selected child root, parent source leaf, child destination
slot, virtual address, PFN roles, and PTE shape. It changes the fresh destination
from \texttt{NORM} to \texttt{PEND}, copies the page in M-mode, and then commits
the child PTE together with the \texttt{PEND}-to-\texttt{PRIV} promotion as one
transaction. Thus the parent and child receive distinct private PFNs even though
Linux retains its native fork bookkeeping.
The agent closes the fork window when the delegated clone returns; closing the
window neither retires nor detaches an admitted child.}

\revision{For a later private COW fault within an attached address space, Linux still
allocates and charges the destination page and performs its ordinary race check
under the PTE lock. It then invokes \texttt{COW\_REPLACE} with the exact slot,
expected old PTE, and candidate destination PTE. OpenSBI derives the source PFN
from the observed protected leaf and validates the fresh destination. It then
changes the destination from \texttt{NORM} to \texttt{PEND}, copies the bytes,
and commits the replacement leaf together with the \texttt{PEND}-to-
\texttt{PRIV} promotion and private-reference updates. \texttt{PEND} is a
transient deny state inside these operations; the implementation maintains no
persistent Linux-controlled copy reservation.}

\para{\revision{Teardown and recycling}}
\revision{\texttt{munmap}, COW replacement, \texttt{exec} teardown, and process
exit reuse the same mediated PTE-clear and exchange sites. OpenSBI retires the
private reference associated with the old protected leaf before Linux completes
its normal RSS, rmap, TLB, and free-page bookkeeping. Protected non-root
pagetable pages are not returned to the Linux buddy allocator: Linux queues
their PFNs in a reclaim list, and \texttt{NACRE\_RECLAIM\_PTP} returns them in
batches to the monitor-managed PTP pool after their leaf mappings have been
retired. The root pagetable page follows a separate lifetime because it was
allocated by Linux and carries the root role; it is retired after the protected
leaves and non-root PTP pages are gone. Finally, \texttt{NACRE\_UNREGISTER}
removes the monitor-side task record and saved execution context, rather than
serving as the address-space cleanup mechanism.}

\subsection{Linux Memory Integration}

\revision{Because the implementation keeps Linux's ordinary memory accounting
and PTE update machinery in place, it must also handle Linux features that
inject or access user mappings outside ordinary anonymous faults. \sysname
handles two such compatibility points in the prototype: executable vDSO code
and S-mode access to private user bytes.}

\para{vDSO integrity}
\revision{Linux still creates the vDSO ABI mapping during \texttt{exec}. During
\sysname initialization, OpenSBI validates the vDSO mapping shape, copies the
accepted image into a monitor-managed protected memory region, and maps that
copy at the expected user vDSO address. The protected process can execute the
usual vDSO stubs, while Linux cannot modify or replace the backing page after
adoption. The associated VVAR data page follows a distinct public-mapping rule:
OpenSBI binds its root, virtual address, PFN, and read-only, non-executable user
permissions, while Linux retains the supervisor mapping used to update the
time data. Thus vDSO/VVAR compatibility does not require reclassifying arbitrary
user pages as public.}

\para{S-mode user-memory access}
Preserving native Linux VM behavior would become difficult to audit if
\sysname added an independent case to every VM and uaccess branch. \sysname
therefore introduces two Linux-side classification markers to keep the
modification surface narrow. \texttt{VM\_NACC\_APP} is a VMA flag for candidate
private-anonymous application mappings in an active \sysname address space.
\texttt{\_PAGE\_NACC} is a software PTE bit that carries private-data intent in
concrete leaf PTEs and across later PTE updates. Together they route fault, COW,
uaccess, and teardown paths toward the protected OpenSBI interfaces. After a
root is attached, OpenSBI defaults ordinary resident user leaves to private
independently of their \texttt{U/W/X} shape, while explicitly recognized
public/system, special, device, and \texttt{PROT\_NONE} mappings remain outside
that class. Linux markers describe candidates and exceptions; they do not
authorize protected ownership.

S-mode access to user memory appears in two forms. Explicit uaccess
uses user virtual addresses and normally reaches shared RISC-V uaccess
primitives that bracket S-mode access by setting and clearing
\texttt{SUM}, the RISC-V status bit that permits S-mode instructions
to access user pages. For a
candidate private range selected by the VMA/PTE markers, those primitives enter
private-data SBI operations instead of opening a raw \texttt{SUM}-enabled access
to the page. OpenSBI resolves the current user address, checks the private leaf
and requested direction, and performs the read, write, copy, or clear in M-mode.
Non-private ranges continue on the normal Linux helper path. \revision{If a
private destination is valid but write-protected for COW, OpenSBI reports a
locked result; Linux clears \texttt{SUM}, enters the ordinary
writable-fault/COW path, and retries after OpenSBI commits the replacement leaf
through the same \texttt{PEND}-to-\texttt{PRIV} path used by private COW.}

Raw S-mode dereference is the case where Linux reaches the same PFN
through a supervisor mapping, such as the direct map, a kmap-like alias, or a
low-level load/store fallback path. \texttt{SUM} is not the relevant permission
gate for these aliases, but the number of legitimate native Linux entry points
that need private user bytes is limited. \sysname handles them as explicit
OpenSBI operations: Linux enters OpenSBI with the intended access and range, and
OpenSBI checks the request against monitor-maintained protected context before
performing the copy or update in M-mode. This differs from designs that route such interactions
through a shared or bounce buffer outside the protected address space. \sysname
keeps the native Linux call path, but Linux does not directly dereference or
modify \texttt{PRIV} PFNs; the data movement is performed only after OpenSBI
validates the protected leaf and the requested operation.
All other accesses to a \texttt{PRIV} PFN are denied by default.

\subsection{Trap and Architectural Support}

The trap implementation follows the fast-path separation in
\autoref{fig:nacre-procedure}. Normal tasks continue to use the ordinary Linux
trap path. A registered \sysname task traps first into the agent through the
entry recorded in \texttt{NACRE\_TWIN\_ENTRY}. The agent saves the protected
context, prepares a Linux-readable trap frame, and transfers control to the
selected Linux handler through \texttt{acall}. Linux returns through a common
\texttt{aret} wrapper. The agent restores the resulting frame after sanitizing
the return \texttt{sstatus} fields described in \S\ref{subsec:lifecycle}; host
service semantics remain untrusted. The \texttt{acall}/\texttt{aret} pair does
not enter M-mode, but a delegated service that accesses private bytes invokes
the typed OpenSBI operations above.

The QEMU architectural model implements the state needed by this path.
\texttt{NACRE\_STATUS} records the registered container and whether the current
confidential-container control flow is executing in Linux or in the agent. The
translation path treats this state as part of the translation context, so
entries filled for one \sysname execution state are not reused as valid hits
after switching to another state. The model also enforces the agent-region
base-and-bound registers. During a pagetable walk, PMP and the agent bounds
protect intermediate PTE accesses; the page-role policy is checked for the
final physical address before QEMU installs the translation. When OpenSBI
changes a page role, QEMU invalidates cached translations for that PFN range
across its virtual CPUs before later accesses proceed under the new role.

\subsection{Prototype Scope}

The QEMU prototype implements the core substrate rather than the complete
target \sysname system. Its implemented paths include monitor-mediated
protected pagetable updates, bitmap/reference-count metadata, private
anonymous-memory transitions, trap delegation through the agent, and the
principal \texttt{fork}/COW/\texttt{exit} paths exercised by our functional
workloads. In particular, the current fork path uses one selected child root
and the monitor-owned \texttt{FORK\_COPY\_INSTALL} transaction; it does not
depend on shared private PFNs or a persistent copy-reservation protocol.

The prototype does not implement owner-tagged cross-container isolation, a
monitor-visible shared-object model, stale-identity protection, or attestation.
Its generic PTE interfaces also lack complete exact-slot and purpose
authorization, and teardown does not yet provide a malicious-host-safe
revoke--flush--clear-before-\texttt{NORM} transaction or complete failure
rollback. Agent entry is validated only on the cooperative prototype path and
has not been established for concurrent multicore execution. These limitations
bound the security claims in \S\ref{sec:security} without changing the
prototype's positive-path behavior.

\section{Security Analysis}
\label{sec:security}

We analyze \sysname against the adversary in \S\ref{sec:threatmodel}. The
target design relies on five invariants:

\begin{packeditemize}
    \item \textbf{I1: Monitor-only protected commits.} Only the monitor can
    commit lower-half entries in a \texttt{ROOT} page, protected non-root page
    tables, or page-role and protected-reference transitions. Linux retains
    authority over upper-half root entries.
    \item \textbf{I2: Linux proposes, but does not authorize.} Linux may run VM
    policy, allocate pages, and construct candidate PTEs, but the monitor treats
    these values as untrusted inputs.
    \item \textbf{I3: Bound execution.} Protected user execution requires a
    fresh monitor binding among a task, \texttt{cid}, address-space root, and
    saved execution context.
    \item \textbf{I4: Domain-bound memory.} A protected page or page-table page
    may be mapped only by its owning \texttt{cid}, unless a live
    monitor-visible grant authorizes the mapping.
    \item \textbf{I5: Authenticated initialization.} Protected execution begins
    only after the monitor accepts the initial image, address space, agent, and
    public-mapping policy.
\end{packeditemize}

\subsection{Memory Isolation}

An adversarial Linux can try to read a private page through its user mapping,
through a direct-map or temporary kernel alias, or by installing a second PTE
to the same physical frame. \sysname places the final data-plane check below
these virtual-address choices. The architectural model checks the current
execution state and final physical-page role: ordinary U-mode and Linux S-mode
cannot access a \texttt{PRIV} page, whereas protected application and trusted
agent execution can. \texttt{PEND} pages are accessible only in M-mode.
Changing an alias therefore does not bypass the page-role check.

Linux also cannot directly rewrite the relevant translation structures.
Protected non-root page tables are S/U-readable for translation but M-mode
writable, and only the monitor commits the lower-half entries of a
\texttt{ROOT} page. Linux retains its normal upper-half kernel mappings. In the
target multi-container design, the monitor further binds every protected page,
page-table page, root, virtual address, and destination PTE slot to an owner and
generation. A wrong-owner mapping or role transition is rejected before either
the PTE or page metadata changes. These checks establish I1, I2, and I4 when the
owner and exact-slot metadata are complete.

\subsection{Lifecycle Transitions}

Protection must hold while a page changes roles. On a private anonymous fault,
Linux may allocate and clear a \texttt{NORM} page because it contains no
application secret. Before the application can access it, the monitor records
the private role and reference and publishes the protected leaf. For an
attached address space, ordinary resident user leaves default to private unless
the monitor recognizes an explicit public, system, special, device, or
\texttt{PROT\_NONE} mapping; Linux's VMA and PTE markers route the request but
do not authorize the exception.

Fork and COW require a stronger copy protocol because the source already
contains confidential bytes. The monitor validates the source mapping and a
fresh destination, changes the destination from \texttt{NORM} to
\texttt{PEND}, and performs the copy in M-mode. A successful transaction
atomically publishes the new PTE and promotes the destination to
\texttt{PRIV}. Fork uses a fresh child PFN, so parent and child never share the
same private PFN; COW replaces one exact protected leaf in its existing address
space. Neither path exposes a \texttt{PEND} page to Linux or relies on a
Linux-controlled reservation.

The reverse target transition first removes protected mappings and stale
translations, then verifies that the last protected reference is gone, clears
the page in trusted mode, and only then publishes it as \texttt{NORM}. A failed
copy or retire operation must publish neither a partial mapping nor a reusable
page containing confidential data. Together these orderings extend I1 across
fault, fork, COW, unmap, \texttt{exec}, and exit.

\subsection{Execution and Trap Mediation}

Linux controls scheduling, but task metadata alone must not authorize
protected execution. The target design admits a user return only when the
monitor-restored task, \texttt{cid}, root, and saved context satisfy I3. A trap
from that execution first reaches the agent. The agent saves the private frame,
constructs the frame delegated to Linux, and uses \texttt{acall} to enter the
selected Linux service. A valid pending \texttt{aret} returns control to the
agent rather than directly to protected U-mode.

The current agent does not establish semantic correctness of arbitrary Linux
results. It restores the Linux-produced frame after clearing \texttt{SIE} and
\texttt{SPP} and reseeding \texttt{sscratch}; applications must still treat
host-service outputs as untrusted. Linux can also return errors, delay the
service, or refuse it entirely. These behaviors do not grant direct access to
the saved private context, but they remain sources of Iago attacks and denial
of service.

\subsection{Mediated Data Access and Sharing}

Some Linux services legitimately consume or produce private user bytes. Rather
than grant Linux a raw mapping, \sysname invokes typed monitor operations. The
monitor resolves the user virtual address, checks the protected leaf and access
direction, and performs the copy or clear in M-mode. This mechanism preserves
the native Linux call path while making private-byte movement explicit. Its
security nevertheless depends on binding each operation to an authorized
purpose and range; a malicious kernel must not be able to synthesize an
otherwise well-formed copy request for arbitrary data.

Not every user mapping is private. The design admits monitor-recognized public
and system mappings, such as immutable vDSO code and a constrained read-only
VVAR page, while defaulting ordinary application leaves to private. Writable
sharing requires a separate protected object. Same-container attachments name
one object and \texttt{cid}; cross-container attachments additionally require a
grant naming the recipient and permissions. Revocation removes every authorized
mapping and stale translation before the object generation can be reused. This
object rule prevents Linux from reclassifying an arbitrary private PFN as
shared.

\subsection{Prototype Security Boundary}

The QEMU prototype provides positive-path evidence for a narrower substrate.
Its final-address policy rejects raw Linux S-mode loads, stores, atomic memory
operations (AMOs), and cache-block operations (CBOs) to \texttt{PRIV} pages and
admits \texttt{PEND} access only from M-mode. OpenSBI writes the protected page
tables and page-role metadata exercised by the functional workloads, while
\texttt{FORK\_COPY\_INSTALL} and \texttt{COW\_REPLACE} combine copy and PTE
publication into one monitor transaction. The agent-first trap path and pending
\texttt{aret} edge are also implemented. \S\ref{sec:evaluation} evaluates these
covered paths.

This evidence does not establish the full malicious-host contract. The current
metadata lacks complete owner/generation and exact-slot authorization; the
Linux scheduler hook and \texttt{pid}/\texttt{cid} registration path are
cooperative; typed data operations are not fully purpose-bound; and generic
failure rollback and trusted clear-before-\texttt{NORM} recycling are
incomplete. Protected shared objects, cross-container grants, multicore-safe
agent entry, authenticated initialization, DMA enforcement, and attestation are
also outside the prototype. Thus the current results support the data-plane
isolation and positive lifecycle paths above, but not yet a complete proof of
confidentiality and integrity against malicious Linux.

\section{Evaluation}
\label{sec:evaluation}

Our evaluation asks four questions: (1) which Linux lifecycle paths does the
prototype execute, (2) what cost appears on common system-call and
communication paths, (3) can the prototype run service workloads, and (4)
which security and scalability claims remain open? We distinguish repeated
measurements from functional and diagnostic tests, and report only evidence
collected with the QEMU prototype.

\subsection{Experimental Scope and Methodology}

The prototype runs in modified QEMU~9.2.0 with a RISC-V Ubuntu~22.04 guest.
The recorded experiments compare the \sysname runtime with a native
\texttt{runc-origin} baseline. For lmbench, both arms use one Docker daemon and select either
the unmodified \texttt{runc-origin} or patched \texttt{runc} with
\texttt{--runtime}. The preserved summaries do not establish the same daemon
control for the other workloads.

We use three evidence classes. Repeated measurements report three executions
per arm. Functional tests establish only that the checked workload and output
complete. Diagnostic tests expose a failure boundary but do not support a
performance claim. Because all experiments run in QEMU, the results
characterize the current software prototype rather than another deployment
environment.

\subsection{RQ1: Which Linux Paths Does the Prototype Execute?}

An eight-path functional regression exercises private-buffer reads and
writes, anonymous memory, \texttt{fork}/COW, file and pipe I/O, \texttt{exec},
teardown, and a shared-memory compatibility path. All eight workloads returned
\texttt{status=ok} and produced their expected outputs. A joint scan of their
QEMU and guest logs found no raw-access denial, typed-copy denial, access
fault, kernel Oops, or panic. This regression belongs to a revision before the
final fork transaction, so it establishes only the covered paths on that
revision.

The implemented fork transaction replaces the earlier multi-step path with a
monitor-owned \texttt{fork\_copy\_install} transaction. Its dedicated
protected-fork test produced the expected output, exited with code zero, and
showed no \sysname rejection or guest crash. The two regressions therefore
provide complementary evidence: the broad regression covers path breadth on
the prior revision, whereas the dedicated test covers the implemented fork
path. A complete broad regression on the same revision remains missing.
Existing Linux Test Project (LTP) records are
dominated by container capability and seccomp policy, and \sysname coverage is
incomplete; we consequently make no Linux-ABI completeness claim from them.

\subsection{RQ2: What Cost Appears on Common Paths?}

Table~\ref{tab:lmbench-preliminary} reports five stable metrics from the
same-daemon lmbench experiment. Each entry is the mean of three runs, and every
\sysname mean is within 3.5\% of the corresponding baseline mean. We omit process, Unix-socket,
file-syscall, and memory-bandwidth cases whose records contain missing values or
large outliers.

\begin{table}[t]
\centering
\scriptsize
\setlength{\tabcolsep}{3pt}
\renewcommand{\arraystretch}{1.12}
\caption{Same-daemon lmbench measurements on QEMU. Values are three-run
means; latency is lower-is-better and bandwidth is higher-is-better.}
\label{tab:lmbench-preliminary}
\begin{tabular}{lrrr}
\toprule
\textbf{Case} & \textbf{Origin} & \textbf{\sysname} & \textbf{Ratio} \\
\midrule
null syscall ($\mu$s) & 4.969 & 4.894 & 0.985 \\
read syscall ($\mu$s) & 8.523 & 8.389 & 0.984 \\
write syscall ($\mu$s) & 5.892 & 5.924 & 1.005 \\
pipe latency ($\mu$s) & 730.357 & 705.012 & 0.965 \\
pipe bandwidth (MB/s) & 47.803 & 47.983 & 1.004 \\
\bottomrule
\end{tabular}
\end{table}

The three repetitions are insufficient to interpret small apparent speedups,
and the pipe-latency samples are mixed. The supported conclusion is narrower:
the retained syscall and pipe means show no large software-path regression in
this QEMU experiment. These aggregate measurements do not attribute cost to
\texttt{acall}/\texttt{aret}, PFN-role checks, protected PTE commits, or TLB
effects.

\subsection{RQ3: Can the Prototype Run Service Workloads?}

For nginx, we transferred 1~GiB at concurrency 16 for each of eight static
object sizes from 4~KiB to 512~KiB, with three runs per size and arm. All 48
runs completed with zero failed requests. Across the eight per-size means, the
\sysname/origin throughput ratio ranges from 0.950 to 1.004. The ratio of the
equally weighted means is 0.9813 (16,557.46 versus 16,873.20~KB/s). Thus, this
QEMU matrix shows that nginx completes while its measured aggregate throughput
is 1.87\% below the baseline; it does not provide a confidence interval over
independent experiment matrices.

Three additional workloads provide functional sanity evidence. Five netperf
configurations completed with zero return codes. SQLite completed the same SQL
workload in both arms and produced identical output hashes. A memcached probe
completed a direct \texttt{set}/\texttt{get} and a one-request memtier
invocation. Because these workloads lack repeated, controlled comparative
measurements, we use them only to demonstrate execution of the tested paths.

\subsection{RQ4: What Remains Open?}

Hackbench identifies a scalability boundary rather than a benchmark result.
With 2,000 workers and five loops, both arms returned zero, but the \sysname
run took 1.50$\times$ longer according to Hackbench's internal timer and
4.59$\times$ longer end to end. Most additional time precedes the internal
timer, suggesting---but not establishing---a cost in worker creation,
registration, or readiness. A larger
\sysname run reached \texttt{CONTAINER\_FULL} and failed to attach a child.
These single-run observations expose capacity and lifecycle work; they do not
establish steady-state scalability.

The current security evidence is likewise positive-path evidence. The broad regression
completed without falling back to raw S-mode access to \texttt{PRIV} pages,
and the dedicated fork test exercised the final monitor-owned copy-and-install
path. We have not run negative tests for wrong owner or \texttt{cid}, stale
task bindings, role or permission changes, failed \texttt{PEND} transactions,
or clear-before-reuse. Cross-container sharing, multicore agent entry, and
attestation also remain unevaluated. The evaluation therefore supports the
feasibility of the exercised single-container paths, not a complete
malicious-host security or compatibility claim.

\begin{table*}[t]
\centering
\scriptsize
\setlength{\tabcolsep}{4pt}
\renewcommand{\arraystretch}{1.12}
\caption{Evidence boundary of the current QEMU prototype.}
\label{tab:claim-experiment}
\begin{tabularx}{\textwidth}{
>{\raggedright\arraybackslash}p{0.20\textwidth}
>{\raggedright\arraybackslash}p{0.28\textwidth}
>{\raggedright\arraybackslash}p{0.22\textwidth}
>{\raggedright\arraybackslash}X}
\toprule
\textbf{Claim} & \textbf{Current evidence} & \textbf{Supported scope} &
\textbf{Evidence still required} \\
\midrule
Linux lifecycle paths & Prior-revision eight-path regression; implemented-fork
dedicated test & Honest execution of the covered fault, uaccess, fork,
\texttt{exec}, and teardown paths & Broad regression on one final revision and
a wider compatibility suite \\
\midrule
Common-path cost & Three-run same-daemon lmbench subset & Five QEMU syscall and
pipe means within 3.5\% & More repetitions, distributions, and mechanism-level
decomposition \\
\midrule
Service execution & Repeated nginx matrix; netperf, SQLite, and memcached
functional checks & Completion of the tested service paths and nginx throughput
in one QEMU matrix & Controlled repeated application experiments \\
\midrule
Protected transitions & Strict positive-path regression and final fork
copy/install test & Exercised single-container transitions & Wrong-domain,
stale-state, failed-transaction, and reuse negative tests \\
\bottomrule
\end{tabularx}
\end{table*}

\section{Discussion and Limitations}
\label{sec:discussion}

\para{Prototype scope}
The current artifact consists of the core \sysname software stack and a QEMU
model of its architectural mechanisms.  It exercises protected private-memory
transitions, monitor-mediated page-table commits, agent-directed trap
delegation, and the main single-container \texttt{fork}/COW/\texttt{exit}
paths.  The results in this paper therefore establish feasibility only for
these modeled paths; they do not validate the target design's complete
multi-container, sharing, multicore, or attestation semantics.

\para{Architectural realization and sPMP}
Region mechanisms such as PMP or sPMP can deny ordinary S-mode accesses to
reserved monitor and agent memory.  They cannot distinguish containers within
host-managed memory, bind translations to a container, or deliver a protected
trap to a peer-privilege agent.  \sysname therefore uses region protection as
a bootstrap boundary, then adds architectural container-identity state,
identity-aware translation checks, and identity-aware event delivery.  The
QEMU model captures only the functional semantics of these mechanisms.

\para{Portability}
The design requires an unforgeable domain identity, protected ownership
metadata, a commit path that the host cannot bypass, and protected event
delivery.  Arm or x86 mechanisms could approximate parts of this interface,
but placing the agent in a separate VM, realm, or secure-world context may
reintroduce address-space switches, additional trusted software, or extra
privilege transitions.  We do not evaluate the security or performance of
such ports.

\para{Untrusted I/O and trusted paths}
\sysname deliberately leaves Linux and its device drivers outside the TCB.
Even when trusted code bounds a delegated buffer and sanitizes architectural
return state, it cannot make a malicious filesystem, network stack, or device
available or correct.
This limitation follows from treating the host system-call interface as an
untrusted service boundary~\cite{checkoway2013iago}.
Confidential storage and network channels therefore require authenticated
encryption to a trusted endpoint, while direct device assignment additionally
requires DMA isolation and a device-specific reset and ownership protocol.  A
verifiable user path could provide stronger assurance for selected devices,
but is orthogonal to the CPU execution boundary studied here~\cite{zhou2012building}.

\section{Related Work}
\label{sec:related}

\para{Enclaves and shielded applications}
Enclave systems protect an application address space from an untrusted OS.
Haven, SCONE, Graphene-SGX, Panoply, SGX-LKL, and Occlum recover different
subsets of OS functionality through library operating systems, shielded
runtimes, or carefully mediated host calls~\cite{baumann2014haven,arnautov2016scone,gramine203255,shinde2017panoply,priebe2019sgxlkl,shen2020occlum}.
Keystone and Penglai provide extensible or scalable enclave abstractions on
RISC-V~\cite{lee2020keystone,feng2021penglai}.  These systems establish strong
application-centric protection, but multi-process and resource-sharing
semantics must be reconstructed inside or coordinated across enclave
boundaries.  \sysname instead makes the Linux container process group the
target protection and lifecycle unit while retaining host-native process and
memory management.

\para{Confidential virtual machines}
AMD SEV-SNP, Intel TDX, and Arm CCA protect guest execution from a hostile
hypervisor or host~\cite{sev2020strengthening,tdx2020,cca2023}.  Kata Containers
and Confidential Containers use this VM boundary to protect containerized
workloads with a guest kernel and supporting services inside the trusted
domain~\cite{randazzo2019kata,confidentialcontainers,coco2026design}.
Firecracker reduces the footprint and startup cost of this virtualization
model for serverless workloads~\cite{agache2020firecracker}.  These approaches
offer mature Linux semantics within the guest, whereas \sysname explores a
container-granularity boundary that lets host Linux retain scheduling and
resource management without placing it in the security TCB.

\para{Container-granularity protection}
TZ-Container, BlackBox, ConMonitor, RContainer, and Fasco protect containers
from an untrusted host using combinations of TrustZone, virtualization,
container monitors, and confidential-computing extensions~\cite{hua2021tzcontainer,van2022blackbox,xu2024conmonitor,zhou2025rcontainer,song2026fasco}.
Arca and ACE further explore lightweight confidential execution for cloud-native
or embedded settings~\cite{lu2026arca,ozga2025ace}.  These systems differ in
their trusted service layer, protection context, and division of lifecycle
responsibility.  \sysname occupies another point in this design space: it
introduces a hardware-recognized container identity and routes protected events
to an isolated peer-privilege agent, while a higher-privilege monitor commits
security-sensitive ownership and mapping transitions.

\para{Protection over an untrusted commodity OS}
Overshadow and InkTag retrofit protected application execution over an
untrusted commodity OS using virtualization and authenticated or trusted
memory-management interfaces~\cite{chen2008overshadow,hofmann2013inktag}.
Dune exposes privileged CPU facilities to applications, illustrating how
hardware virtualization can support protection mechanisms outside the kernel
without replacing the kernel's service role~\cite{belay2012dune}.  Iago attacks
show that protecting memory alone is insufficient when an adversarial OS can
manipulate system-call results~\cite{checkoway2013iago}.  \sysname shares the
principle that the host may provide policy and services but must not authorize
protected transitions; it specializes that separation to native container
identity, trap delivery, and Linux lifecycle operations.

\para{Paravirtualized secure containers}
ParaCell uses memory protection keys (MPK) for intra-address-space isolation
and intent-driven memory management to reduce user/kernel transition cost and
improve memory elasticity for paravirtualized secure
containers~\cite{wu2026paracell}.  Its host Linux
acts as the virtual-machine monitor and remains responsible for container
scheduling, interrupt forwarding, device I/O, and host mappings.  ParaCell
therefore addresses efficiency under a trusted-host model, whereas \sysname
separates host resource management from security authority under a compromised
host.

\para{Elastic and confidential serverless execution}
Reusable enclaves, plug-in enclaves, and Split Containers reduce initialization
or memory cost for confidential serverless functions~\cite{zhao2023reusable,li2021confidential,shi2025serverless}.
Other systems accelerate function creation through optimized containers,
snapshots, or language/runtime isolation~\cite{oakes2018sock,akkus2018sand,cadden2020seuss,du2020catalyzer,shillaker2020faasm,ao2022faasnap}.
These techniques are complementary to \sysname's focus: they optimize how an
execution environment is created or reused, while \sysname asks how a dynamic
Linux container can remain the hardware-recognized protection unit when the
host kernel is untrusted.

\section{Conclusion}

We presented \sysname, which makes a Linux container process group the
protection unit while leaving scheduling and resource management to an
untrusted host.  A hardware-recognized container identity and an isolated
S-mode agent keep frequent interactions on the native Linux path, while an
M-mode monitor commits protected mappings and lifecycle transitions.  Our QEMU
prototype exercises the core single-container private-memory and trap paths,
including \texttt{exec}, demand paging, \texttt{fork}, COW, and teardown.  In
the recorded experiments, the three-run means of five lmbench syscall and pipe
metrics were within 3.5\% of the same-daemon \texttt{runc-origin} baseline, and the
\sysname/baseline ratio of equally weighted nginx mean throughputs across eight
object sizes was 0.981.  These results support the feasibility of separating
host resource management from security authority on the modeled data and
control paths.  The cooperative identity and lifecycle checks that remain in
the prototype delimit, rather than complete, its malicious-host security claim.

\bibliographystyle{IEEEtranS}
\bibliography{ref}

\end{CJK*}
\end{document}